\documentclass[twocolumn,showpacs,amsmath,amssymb]{revtex4-1}

\usepackage{graphicx}
\usepackage{dcolumn}
\usepackage{bm}
\usepackage{url}
\usepackage{amssymb}
\usepackage{wasysym}
\usepackage[stable]{footmisc}
\usepackage{slashbox}
\usepackage{longtable}
\usepackage{tabularx}
\usepackage{float}
\usepackage{mathptmx}
\usepackage{relsize}
\begin{document}
\title[Supercond. Sci. Technol.]{Localized High Frequency Electrodynamic Behavior of Optimally-doped Ba(Fe$_{1-x}$Co$_x$)$_2$As$_2$ Single Crystal films}

\author{Tamin Tai$^{1,2}$, Behnood G. Ghamsari$^{2}$, J. H. Kang$^{3}$, S. Lee$^{3}$, C. B. Eom$^{3}$, Steven M. Anlage$^{1,2}$}
\address{$^{1}$ Electrical and Computer Engineering, University of Maryland, College Park, Maryland 20742-3285, USA}
\address{$^{2}$ Department of Physics, Center for Nanophysics
and Advanced Materials (CNAM), University of Maryland, College Park, Maryland  20742-4111, USA}
\address{$^{3}$ Department of Materials Science and Engineering, University of Wisconsin-Madison, Madison, WI 53706, USA}

\vspace{8pt}
\newcommand{\squeezeup}{\vspace{-2.0mm}}

\begin{abstract}
Localized high frequency (several GHz) electrodynamic properties of a high quality epitaxial, single-crystal Iron-Pnicitde Ba(Fe$_{1-x}$Co$_x$)$_2$As$_2$ thin film near optimal doping (x=0.08) are measured under a localized and strong RF magnetic field, created by a near-field microwave microscope. Two reflection electrodynamic measurements, including linear and the third harmonic responses, are performed to understand the electromagnetic properties of Ba(Fe$_{1-x}$Co$_x$)$_2$As$_2$. Our measurement results show that Ba(Fe$_{1-x}$Co$_x$)$_2$As$_2$ has have a wide superconducting transition width and may have a multi-gap nature. In addition, based on the $1/T^2$ dependence of the third harmonics signal at lower temperature, Ba(Fe$_{1-x}$Co$_x$)$_2$As$_2$ shows the possibility of nodal behavior.

\end{abstract}



\maketitle

\section{Introduction}
The superconducting properties of iron-based superconductors have been widely discussed recently \cite{Paglione}\cite{Gurevich_2011}\cite{Terashima}. Because of their high critical temperature ($T_c$) and high upper critical field, many ideas have been proposed for potential applications in superconducting wires for high-field accelerator magnets and many varieties of superconductive devices. These applications all require a high quality iron-based superconducting material. However it is not easy to grow a perfect single crystal of these superconductors. The electron doped iron-pnictide Ba(Fe$_{1-x}$Co$_x$)$_2$As$_2$ (Ba-122 family) can be used to prepare high quality single crystal films by deposition of a SrTiO$_3$ (STO) or BaTiO$_3$ (BTO) template on LSAT or LAO perovskite substrates \cite{S. Lee_2010}\cite{S. Lee_2011}. Although its application is still constrained by the requirement of a buffer layer, it appears that a large critical current density can be achieved. Questions have arisen about whether it has a single \cite{Chen}\cite{Dressel} or multiple energy gaps \cite{Terashima}\cite{Perucchi}, without nodes \cite{Terashima}\cite{Yong}\cite{Xiaohang} or with nodes \cite{Prozorov}, and whether or not it has isotropic \cite{Yong}\cite{Xiaohang} or anisotropic \cite{Prozorov} gaps. This material family is still of intense interest and controversy.

Because of the success of preparing high quality single crystal epitaxial films of Ba(Fe$_{1-x}$Co$_x$)$_2$As$_2$ and on some other Ba-122 families, these films are particularly suitable for the study of the iron-pnictide superconducting gap nature. From many theoretical predictions and experimental measurements, many scenarios for its gap nature are proposed and have created a great deal of debate. Some theoretical predictions have been proposed with $s^{\pm}$ symmetry of the gap on the Fermi surfaces, consistent with the ARPES measurement which shows isotropic gaps without nodes \cite{Terashima}. Temperature dependent penetration depth measurement with power law behavior, $\Delta \lambda(T) \propto T^{n}$ with $n>2$, have been widely discussed but authors have different interpretations for the existence of either anisotropic gap with nodes \cite{Prozorov} or an isotropic nodeless gap \cite{Yong}. Meanwhile, Raman scattering experiments indicate a gap with accidental nodes, which may be lifted by doping and/or impurity scattering in iron arsenides Ba(Fe$_{1-x}$Co$_x$)$_2$As$_2$ \cite{Fisher}.
Optical measurement at terahertz frequency provides the results of either two optical gap superconductivity \cite{Perucchi} or just one gap \cite{Dressel}. One reason that different, or even the same, approaches yield different results may be due to the impurity effects and surface inhomogeneities of the tested single crystal \cite{Moler2009}. Therefore, a localized measurement technique on these Fe-based superconductors should be applied to further illuminate the nature of the gap structure.

Scanning superconductor quantum interference device (SQUID) susceptometry has been used in localized measurement on many iron pnictide superconductors to identify inhomogeneities of superconductivity, correlate these properties to the surface microstructure and to explore the gap nature \cite{Moler2009}. This technique has been used to observe the existence of twin boundaries in the underdoped BaFe$_{1-x}$Co$_x$As$_2$ films, along with an enhancement of the local superfluid density \cite{Moler2010}. In addition, the relation between microstructure, grain boundaries and its critical field attracts many material science researchers to study the iron pnictide materials \cite{Larbalestier_2011}.  For example, the Ba(Fe$_{1-x}$Co$_x$)$_2$As$_2$ epitaxial films shows very high upper critical field and anisotropic physical properties due to randomly distributed BaFeO$_2$ nanorods, which yield very strong vortex pinning in the matrix of the Co-doped Ba-122 thin film \cite{Tarantini}\cite{Zhang}.

In this study, we use our novel near field magnetic-field microwave microscope to detect the localized electromagnetic responses on Ba(Fe$_{1-x}$Co$_x$)$_2$As$_2$ single crystal films with x=0.08. Compared to the scanning SQUID susceptometry, our microwave microscope \cite{Anlage_2001}\cite{Anlage_2007} can create an intense and localized magnetic field on the scale of $\sim$ 200 $mT$ and sub-micron resolution at GHz frequencies by utilizing a hard drive magnetic write head as a scanning probe on superconductors \cite{Tai2011}\cite{Tai2012}\cite{Tai2013}\cite{Tai2014_JAP}\cite{Tai2014_APL}\cite{Tai2013_dissertation}. This method is good for measuring harmonic generation below the superconducting transition temperature, the surface homogeneity of superconductors, and the intrinsic behavior of the superconducting order parameter. In this paper, linear response and third harmonic nonlinear response measurements on a Ba(Fe$_{1-x}$Co$_x$)$_2$As$_2$ film will be addressed and compared to the results on conventional Nb \cite{Tai2014_JAP} \cite{Tai2014_APL} superconductors. Understanding the reflected linear and nonlinear mechanisms including intrinsic and extrinsic nonlinearity in this Ba(Fe$_{1-x}$Co$_x$)$_2$As$_2$ film will help illuminate the nature of iron pnictide materials.

\section{Experiment}

\begin{figure}
\centering
\includegraphics [width=3.3 in]{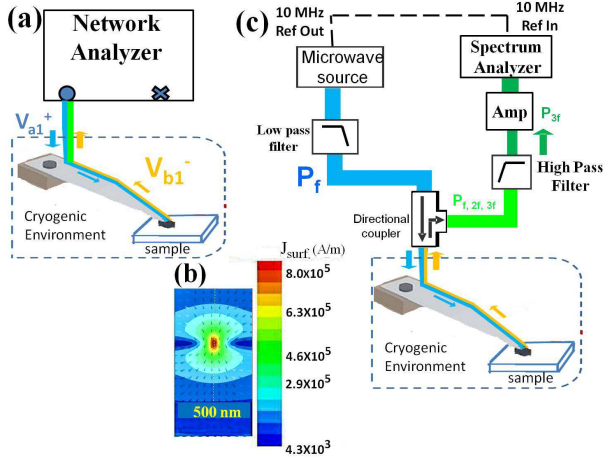}
\caption{(a) Schematic diagram of the linear response measurement, $S_{11}$,
 performed as a function of temperature with the network analyzer. Note that only one port (marked with a ${\bullet}$ symbol) is used in the measurement. (b) A simulated surface current density distribution ($J_{surf}$) on the sample surface created by the magnetic write head probe at the driving frequency is shown, assuming the probe height is 200 $nm$ away from the sample surface \cite{Tai2013}. (c) Schematic set up of the third harmonic measurement in nonlinear microwave microscopy. Note that the third harmonic signal (P$_{3f}$) is selectively filtered by the microwave circuit outside the cryogenic environment. }
\label{setup}
\squeezeup
\end{figure}

Ba(Fe$_{1-x}$Co$_x$)$_2$As$_2$ with x=0.08 (thickness:200 nm) is grown by pulsed laser deposition on a (001) oriented (La,Sr)(Al,Ta)O$_3$ (LSAT) substrate. A 20 nm thick BaTiO$_3$ (BTO) film is deposited as a template layer before the deposition of Ba(Fe$_{1-x}$Co$_x$)$_2$As$_2$ films. Details of the growth conditions can be found in references \cite{S. Lee_2010}\cite{S. Lee_2011}. After growth, a 20 nm thick layer of Pt is capped on the top of Ba(Fe$_{1-x}$Co$_x$)$_2$As$_2$ to protect the film from degradation. The resulting film has T$_c=18.1$ $K$, characterized by the temperature dependent dc resistivity $\rho(T)$ measurement.

We stimulate this Ba(Fe$_{1-x}$Co$_x$)$_2$As$_2$ single crystal film locally with a strong RF magnetic field from a near-field microwave probe and then measure the linear-response reflected signal \cite{Tai2014_JAP} and the reflected third harmonic signal generated by the Ba(Fe$_{1-x}$Co$_x$)$_2$As$_2$ film. Fig. \ref{setup}(a) shows a schematic diagram of the linear response measurement. A single frequency fundamental tone ($V_{a1}^{+}$) is sent out from port 1 of the network analyzer down to the probe on the Ba(Fe$_{1-x}$Co$_x$)$_2$As$_2$ thin film.  A reflected signal ($V_{b1}^{-}$) at the same frequency is collected at the same port and then a ratioed measurement of the complex $V_{b1}^{-}(T)/V_{a1}^{+}(T)$, defined as S$_{11}$ on the Ba(Fe$_{1-x}$Co$_x$)$_2$As$_2$ films, is performed at different
temperatures.  Note that this microscope utilizes a magnetic writer with 100 nm wide magnetic gap made by Seagate for longitudinal magnetic recording technology. The magnetic writer approaches the surface of the superconductor to a distance estimated to be on the order of 0.2 $\mu m$ $\sim$ 1 $\mu m$, which can be approximately judged by the resonant frequency of the probe assembly and by our previous measurements on many superconducting thin films and High Frequency Structure Simulator (HFSS) field strength and configuration modeling. Fig. \ref{setup}(b) shows the simulated HFSS result of the surface current density distribution produced by the magnetic write head probe assembly on the top of a perfect conductor \cite{Tai2013}. The $J_{surf}$ scale bar and arrows indicate the magnitude and
direction of the screening current, respectively, in the first half of the RF cycle.
In this simulation, we assume the yoke in the magnetic writer is made of ferrite. The yoke is excited by a 50 $mA$ RF current and the separation between the probe and the sample
is 200 nm. This creates a maximum surface current density $J_{surf}$=$8*10^5$ $A/m$.

Fig. \ref{setup}(c) schematically shows the nonlinear response setup. In this case, we are gathering the localized third harmonic response generated by the superconductor. Compared to the linear response setup, the microwave circuit outside the cryogenic environment is changed to selectively filter the $P_{3f}$ signal, and the microwave circuit inside the cryogenic environment remains the same. The fundamental principle of operation of the nonlinear microwave circuit and the response from superconductors can be found in our previous work \cite{Tai2011} \cite{Tai2012}\cite{Tai2014_APL}.

\section{Results and Discussion}

\begin{figure}[!t]
\centering
\includegraphics [width=1.5 in]{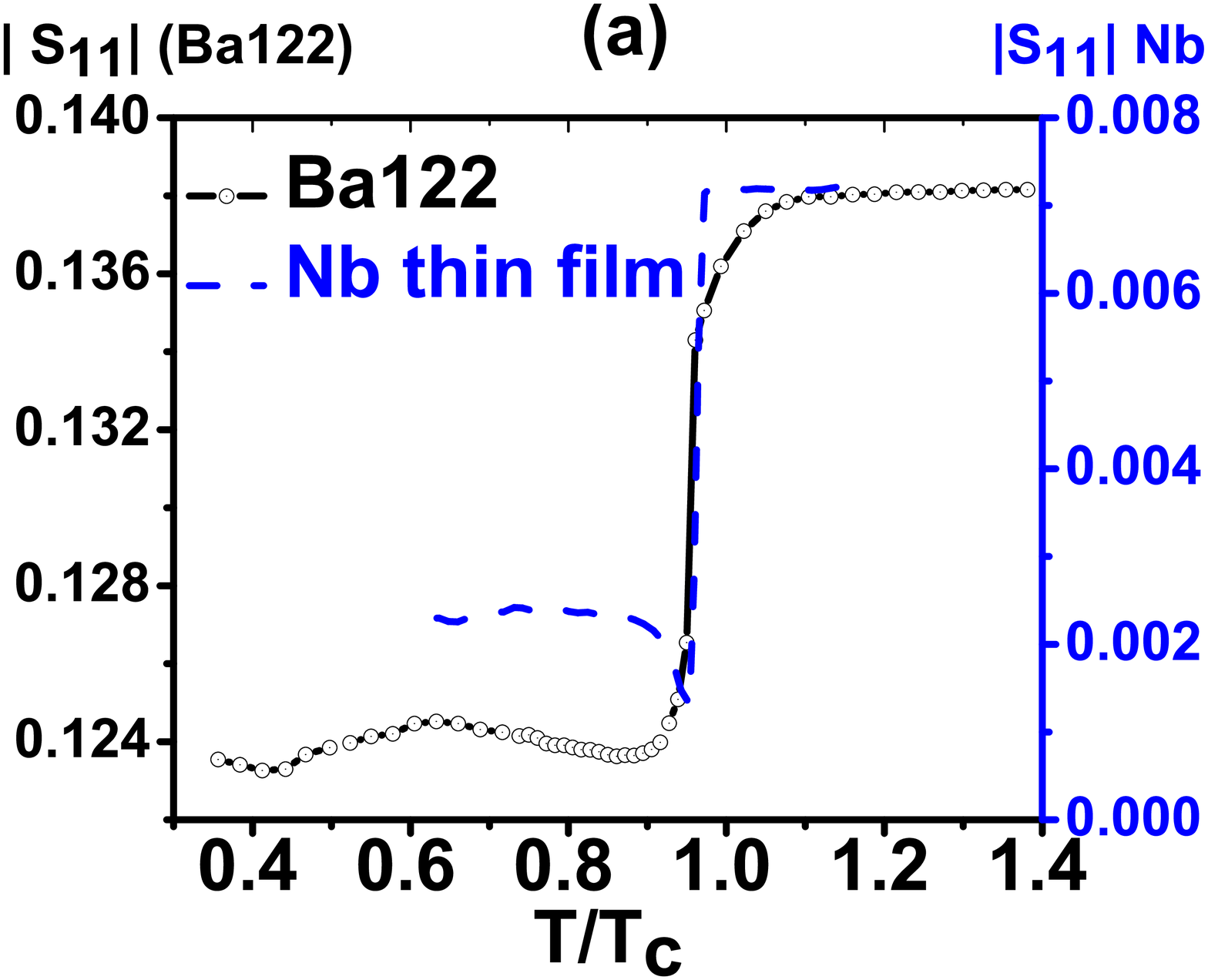}
\includegraphics [width=1.5 in]{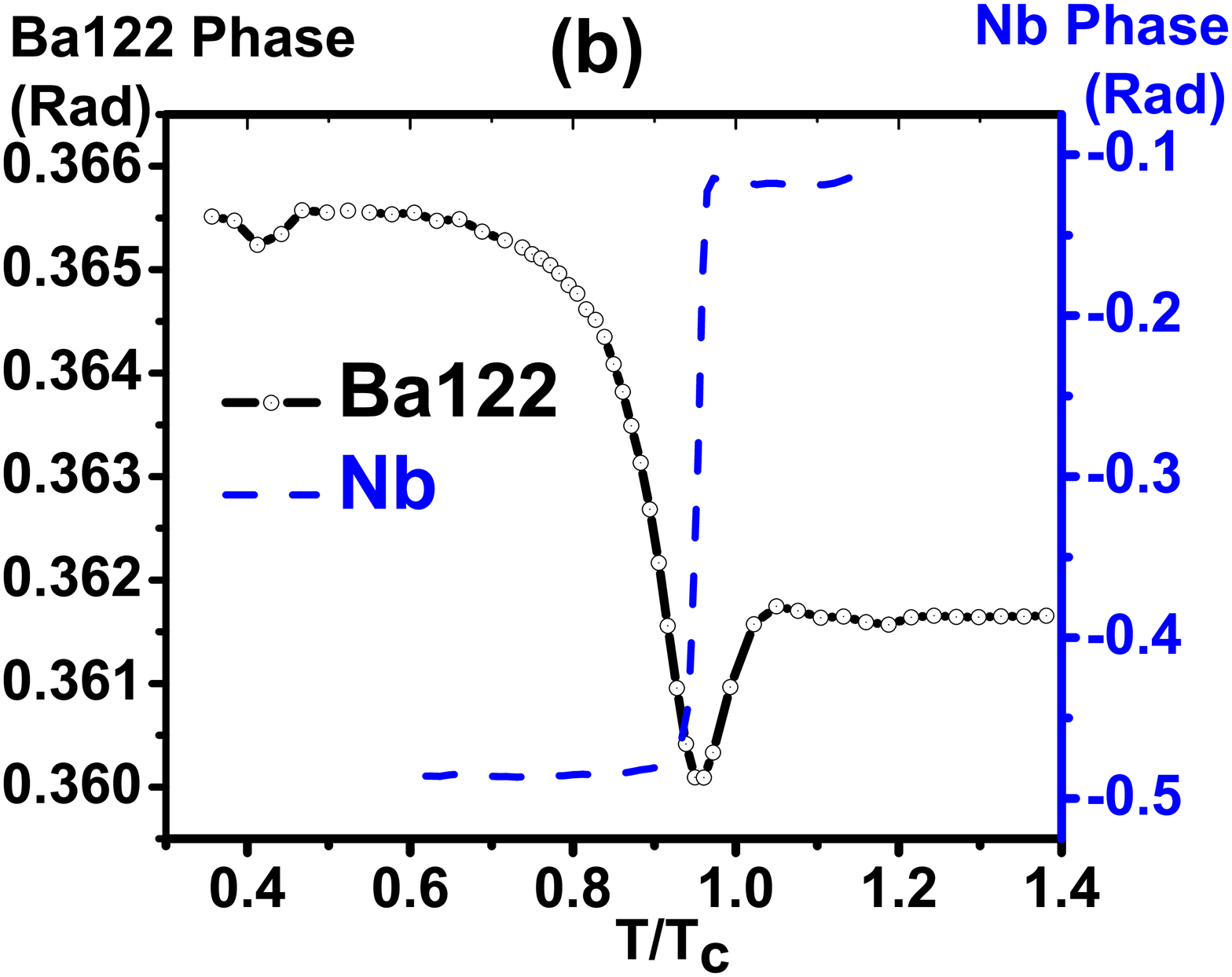}
\caption{The temperature dependent linear response $S_{11}$ of the Ba(Fe$_{1-x}$Co$_x$)$_2$As$_2$ film measured with the near-field microwave microscope. Both (a) amplitude and (b) phase show a transition at normalized $T/T_c=1$. The same measurement for (a) amplitude and (b) phase of S$_{11}$ is also done on a Nb thin film as shown in blue dash lines for comparison. Note the T$_c$ for the Ba(Fe$_{1-x}$Co$_x$)$_2$As$_2$ film and the Nb film are 18.1 K and 8.3 K, respectively. }
\label{S11}
\squeezeup
\end{figure}

Localized linear responses (S$_{11}$) measurement is a useful tool to identify the film transition temperature T$_c$ in a localized area and further estimate its magnetic penetration depth ($\lambda$). The black dots (connected by a solid black line) in Fig. \ref{S11} shows the temperature dependent S$_{11}$ amplitude (Fig. \ref{S11}(a)) and its phase (Fig. \ref{S11}(b)) of the Ba(Fe$_{1-x}$Co$_x$)$_2$As$_2$ film. For comparison, a conventional Nb thin film with T$_c$=8.3 K is also plotted in the blue dash line. For both measurements, the incident power levels are in the linear response regime. A sharp change of amplitude and phase in $S_{11}$ occurs near T/T$_c$=1 for both superconducting films, indicating the individual transition of each film. This sharp change indicates that from the normal state to the superconducting state, the surface impedance of the superconducting films suddenly changes, which results in a change of the reflected voltage ($V_{b1}^{-}$). Quantitative interpretation of this sharp change around T$_c$ has been modeled by combining a magnetic circuit (magnetic write head inductively coupled to the sample)
and transmission line (microwave circuit) \cite{Tai2014_JAP}. However, in the Ba(Fe$_{1-x}$Co$_x$)$_2$As$_2$ measurement, slightly above T$_c$, both the amplitude and phase of Ba(Fe$_{1-x}$Co$_x$)$_2$As$_2$ films show a fluctuation tail before going into the normal state (compare to the Nb result). This tail most likely indicates a wider distribution of T$_c$ values for the Ba(Fe$_{1-x}$Co$_x$)$_2$As$_2$ film.

The high frequency microwave nonlinear response measurement may be useful to figure out the intrinsic multi-gap nature of Iron Pnictide Ba(Fe$_{1-x}$Co$_x$)$_2$As$_2$ films. In addition, the nonlinear measurement is also very sensitive to the surface defects, for example due to extrinsic defects which generate additional channels of dissipation and reactance. In order to measure the nonlinear response of this Ba(Fe$_{1-x}$Co$_x$)$_2$As$_2$ film, we change our microwave circuit to that in Fig. \ref{setup}(b) and probe the Ba(Fe$_{1-x}$Co$_x$)$_2$As$_2$ single crystal film surface at a fixed position.

\begin{figure}
\center
\includegraphics[height=2.3in]{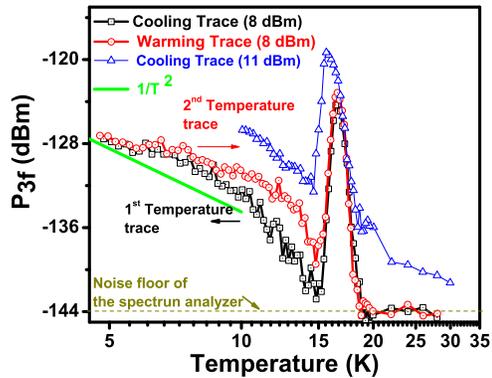}
\caption{Temperature dependence of the third harmonic response (log-log plot) for a
200 nm thick Ba(Fe$_{1-x}$Co$_x$)$_2$As$_2$ film under 5.1 GHz microwave excitation. The square and circle shapes with inner-centered dot points are experimental $P_{3f}$ data under 8 dBm excitation. The data with blue triangular shape is taken under 11 dBm excitation. The green solid line shows a $1/T^2$ trend to guide the eye.} \label{Ba122P3f}
\squeezeup
\end{figure}

Fig. \ref{Ba122P3f} shows a representative result of temperature dependent third harmonic power $P_{3f}(T)$ under 5.1 GHz localized microwave excitation. Excitations at different power are performed to clarify the relation of $P_{3f}(T)$ at low (8 dBm) and high excitation power (11 dBm). The criteria of low power or high power excitation is judged by whether the probe nonlinearity is excited above the noise floor of the spectrum analyzer or not \cite{Tai2011}. Peaks in $P_{3f}(T)$ near $T_{c}$ can be interpreted as the intrinsic nonlinearity from the rf current-induced modulation of the superconducting order parameter near $T_c$ due to the decrease of superfluid density and the associated divergence of the penetration depth \cite{Tai2011}\cite{Lee}\cite{Mircea}. Comparing to the low power excitation and high power excitation $P_{3f}(T)$ curve, one finds the positions of these peaks slightly shift toward lower temperature at higher excitation power due to localized heating by microwave power. In addition, one finds both peaks (microwave $T_c$) are slightly lower than the ``dc zero-resistance $T_c$". The lower microwave $T_c$ implies microwave measurements generally respond to lower dissipation levels, making them more sensitive than the DC resistivity measurements.

\begin{figure}
    \centering
    \includegraphics[width=2.0in]{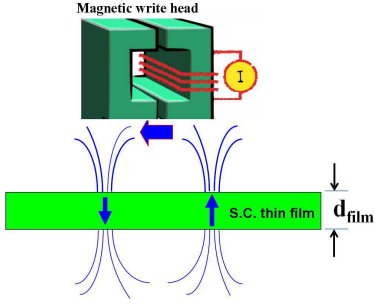}
    \caption{Schematic illustration of the magnetic flux coming from the yoke of the magnetic write head to the superconducting thin film. This situation is analogous to a magnetic dipole on top of the superconducting thin film (horizontal blue arrow). A vortex and antivortex perpendicular to the film tend to form if the film thickness $d_{film} \ll \lambda$.  Figure not to scale.}
    \label{yoke}
    \squeezeup
\end{figure}

In addition, while T $<$ 15 K, $P_{3f}(T)$ increases monotonically with decreasing temperature. Although this observed temperature dependence of P$_{3f}(T)$ is reminiscent of that arising from Josephson weak links \cite{Jeffries} or Josephson vortices in a large Josephson junction \cite{Lee2}, weak link nonlinearity should not occur in the epitaxial single crystal structure. From 15 K to 6 K at 8 dBm excitation, one can find the cooling down nonlinear trace and warming up nonlinear re-trace shows a hysteretic behavior \cite{L. Ji}. This implies that extrinsic nonlinearity due to oscillation and motion of trapped vortices \cite{Gurevich2008} would be one of the possible nonlinear mechanisms at this intermediate temperature regime. Because the thickness of this film is always smaller than the penetration depth, the magnetic flux coming out from the magnetic gap of the writer probe must go through the film and forms a vortex and antivortex pair perpendicular to the film. This situation is analogous to having a parallel magnetic dipole on top of the superconducting thin film (horizontal blue arrow in Fig. \ref{yoke}). A vortex and an antivortex nucleate perpendicular to the film and will move under the influence of the RF screening currents in the film.  One can model this situation with an equivalent point magnetic dipole that is horizontally-oriented and placed above the superconducting thin film. Once the vortices are inside the film, pinning can occur by the randomly distributed BaFeO$_2$ nanorods in the matrix of the Co-doped Ba-122 thin film. The pinned vortices will oscillate under the influence of the fundamental RF currents and generate harmonic response \cite{L. Ji}\cite{Golosovsky}\cite{E. B. Sonin}\cite{M. W. Coffey}. Hence the creation, motion and pinning of perpendicular vortex and antivortex pairs will generate high order harmonic response in the intermediate temperature region.

\begin{figure}
\centering
\includegraphics[width=2.4in]{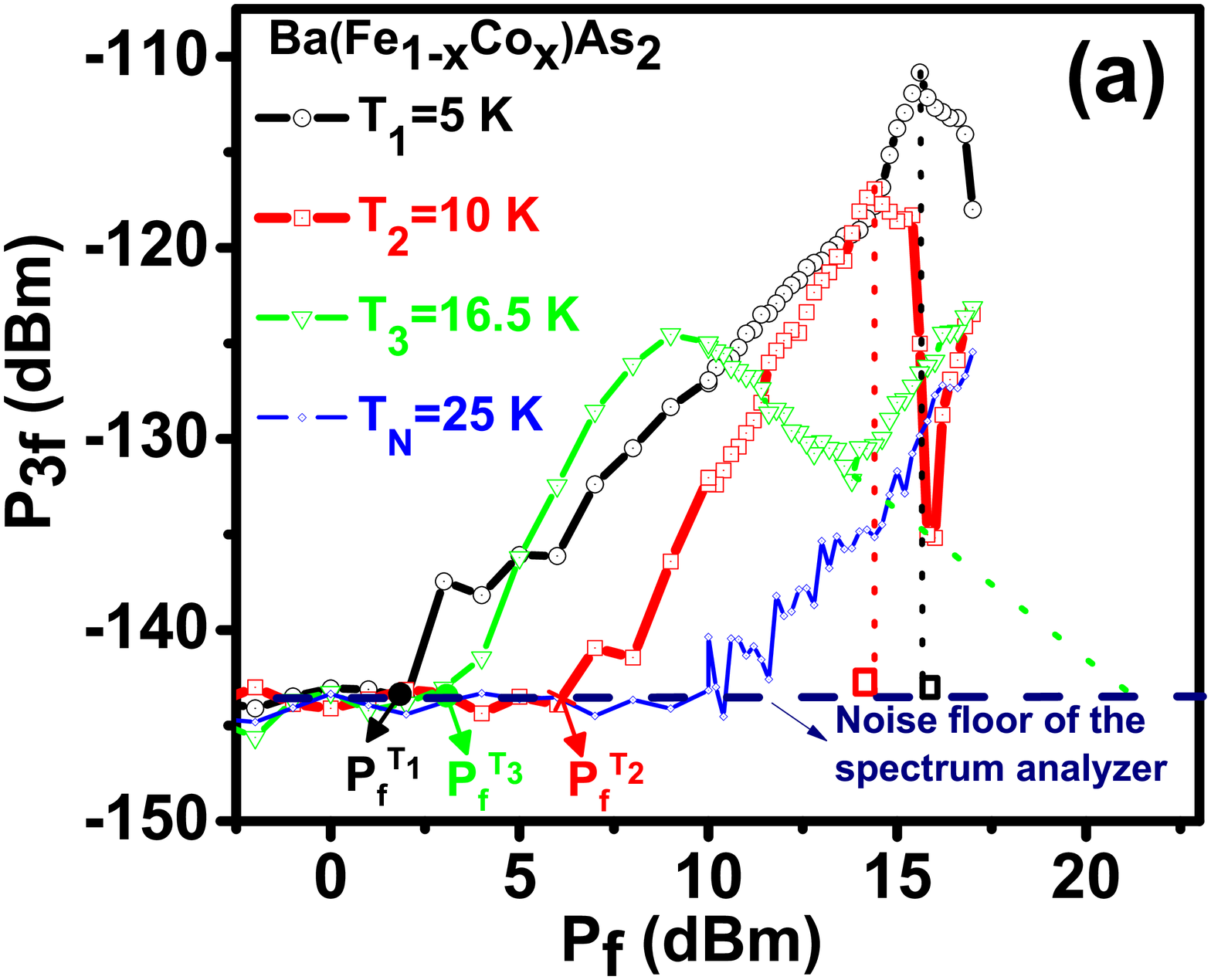}
\includegraphics[width=2.5in]{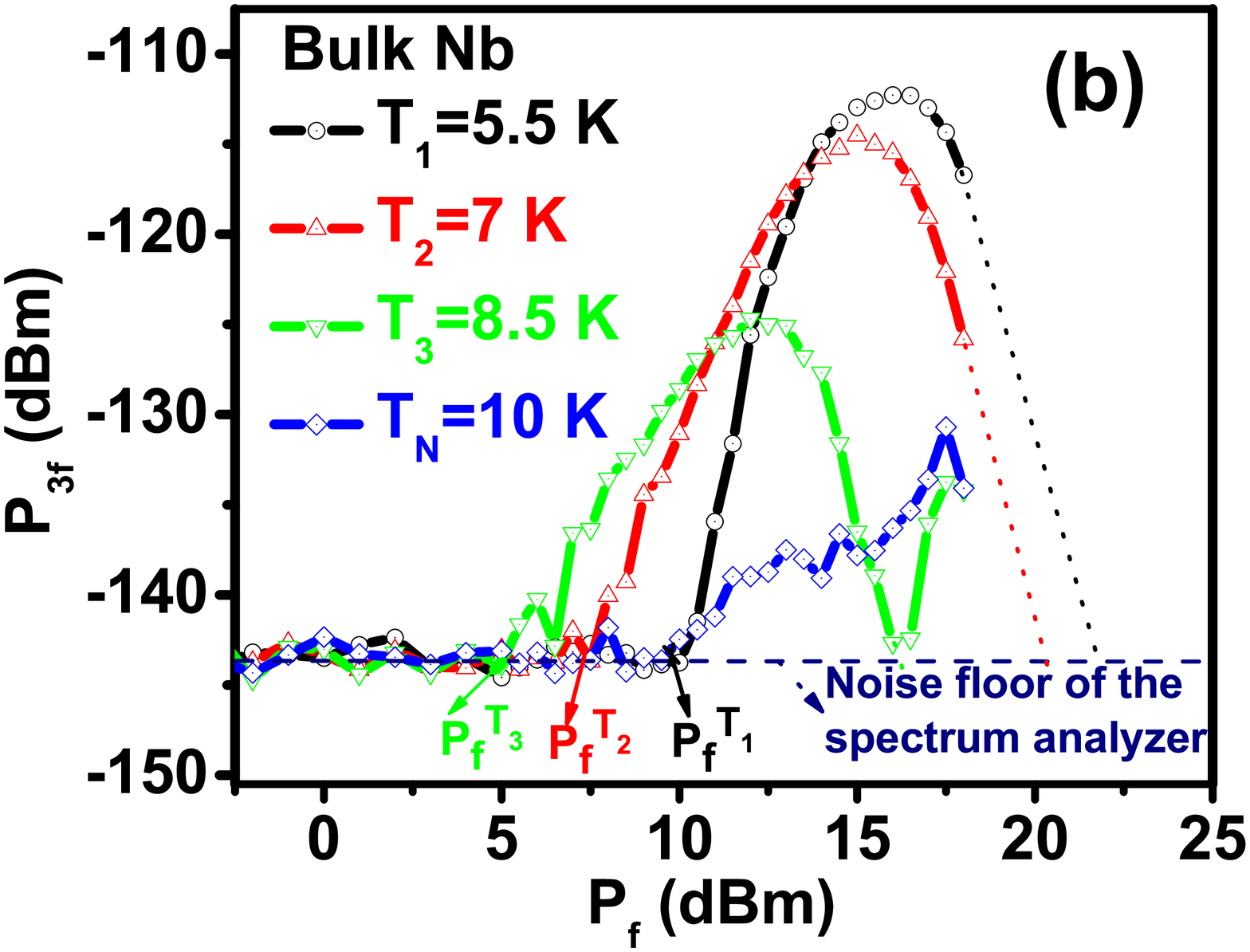}
\begin{quote}
\caption{(a) Dependence of $P_{3f}$ on $P_f$ for a 200 nm thick Ba(Fe$_{1-x}$Co$_x$)$_2$As$_2$ film under 5.1 GHz local microwave excitation. The vertical dashed lines for $T_1=5$ K and $T_2=10$ K indicate the turnover point. Another dashed line on the $T_3$=16.5 K curve indicates the extrapolation of $P_{3f}$ on $P_f$ through the probe nonlinearity, $P^{probe}_{3f}$, which is taken at $T_N$=25 K above the $T_c$ of the Ba(Fe$_{1-x}$Co$_x$)$_2$As$_2$ film. The horizontal dashed line indicates the noise floor ($\sim$ -143 dBm) of the spectrum analyzer.  (b) Power dependence of $P_{3f}$ on $P_f$ for a bulk Nb sample under 5.36 GHz local microwave excitation \cite{Tai2014_APL}. The dashed line in each curve indicates the extrapolation of $P_{3f}$ on $P_f$ through the $P^{probe}_{3f}$ to the noise floor of the spectrum analyzer. The $T_c$ of this bulk Nb is 9.2 K and $P^{probe}_{3f}$ is taken at $T_N$=10 K.}
\label{P3fPf}
\squeezeup
\end{quote}
\end{figure}

The nodal or non-nodal gap features in Ba(Fe$_{1-x}$Co$_x$)$_2$As$_2$ can be judged by temperature-dependent P$_{3f}$(T) at thermal energy below the energy of the smallest gap.
At temperature T $<$ 6 K at 8 dBm excitation power shown in Fig. \ref{Ba122P3f}, there is no hysteresis behavior. This implies that extrinsic nonlinearity  due to vortices does not dominate the nonlinear response because of the increase of pinning force due to the increase of critical current at lower temperature. In addition, the probe nonlinearity, P$_{3f}^{probe}$ at 8 dBm is under the noise floor of the spectrum analyzer, and would not affect the measured P$_{3f}$ (which will be discussed in Fig. \ref{P3fPf}). However, one finds the variation of low temperature dependent P$_{3f}$(T) follows the relation:
\begin{equation}
P_{3f}(T) \propto \frac{1}{ T^2}
\squeezeup
\end{equation}
The green solid line in Fig. \ref{Ba122P3f} is a $1/ T^2$ dependence in the temperature range of 4 K  $\sim$ 10 K, to guide the eye. This $1/ T^2$ behavior in Ba(Fe$_{1-x}$Co$_x$)$_2$As$_2$ is the same as that found in the high-quality d-wave YBa$_2$Cu$_3$O$_7$ (YBCO) superconductor for the intrinsic nonlinear Meissner Effect at low temperature \cite{oates}\cite{Agassi}, and also the same as the claim from Agassi, Oates and Moeckly on the observation of Mg$B_2$ line nodes \cite{oates_2010}.  Therefore, one possible interpretation is that Ba(Fe$_{1-x}$Co$_x$)$_2$As$_2$ has line nodes in the superconducting gap, similar to that observed in the d-wave YBCO superconductor. Another possibility is the nonlinearity coming from transport through normal metal-superconducting (NS) junctions \cite{Lesovik}. Note that the Ba(Fe$_{1-x}$Co$_x$)$_2$As$_2$ film is capped with a 20 nm thick Pt layer. Therefore an NS interface will produce proximity-induced subgap states which are probed by the RF field. In addition, if the interface of the NS junction has impurities or vacancy defects, the normal metal-insulator-superconducting (NIS) junctions may also be formed. Then localized surface states due to the NIS junctions in the multi-band superconductor can also produce significant nonlinearity even in the absence of nodes. We are not aware of quantitative analysis of temperature dependent nonlinearity on NS and NIS junctions of Ba-122 families.

In order to understand the intermediate and low temperature nonlinear mechanisms, measuring the dependence of nonlinearity on the fundamental tone power is one of the approaches. Fig. \ref{P3fPf}(a) shows the dependence of $P_{3f}$ on $P_f$ at a fixed position and some selected temperatures for this 200 nm thick Ba(Fe$_{1-x}$Co$_x$)$_2$As$_2$ film.  In the normal state (T $>$ 18.1 K), the measured nonlinearity comes from the probe itself because the magnetic write head is made of ferrite which generates background nonlinearity \cite{Tai2011}. Note that the probe third harmonic response (P$_{3f}^{probe}$), which also depends on the probe height, only becomes measurable at high excitation powers (above 10 dBm). For measurement below
$T_c$, all curves (at temperature T$_1$,T$_2$ and T$_3$) show a sharp $P_{3f}$
onset from the noise floor of the spectrum analyzer. After the onset, the nonlinearity continues to increase with fundamental power until a turnover point. After the turnover point, the nonlinearity
goes down until it approaches the curve of probe nonlinearity, $P^{probe}_{3f}$ (data taken at $T=T_N$). After that point, the measured nonlinearity oscillates around the curve of probe nonlinearity as the two contributions interfere constructively and destructively. Fig. \ref{P3fPf}(b) is the same power dependent measurement of $P_{3f}$ on $P_f$ but on a bulk Nb superconductor for comparison. All curves also have a sharp $P_{3f}$
onset from the noise floor of the spectrum analyzer, which is -142.5 dBm in this measurement. Nonlinearity keeps growing before a turnover point and then drops gradually to follow the probe nonlinearity. Each curve in Fig. \ref{P3fPf}(a) and Fig. \ref{P3fPf}(b) looks similar but there are many differences upon closer examination.

First, the relation of temperature dependent onset is different. In the bulk Nb measurement, the relation of onset power for each temperature is $P_f^{T_3}$ $<$ $P_f^{T_2}$ $<$ $P_f^{T_1}$. Therefore for conventional superconductors such as Nb, the onset of nonlinearity requires higher excitation power at lower temperatures because the sample remains in the Meissner state and because of the larger critical field ($\sim$H$_{c1}$) at lower temperature. However, in the Ba(Fe$_{1-x}$Co$_x$)$_2$As$_2$ measurement, one can clearly see the relation of onset power is the opposite, $P_f^{T_2}$ $>$ $P_f^{T_1}$ and $P_f^{T_3}$ $>$ $P_f^{T_1}$. This means nonlinearity can be easily excited at lower temperature (T=5 K), the same behavior observed on a two gap high quality MgB$_2$ film \cite{Tai2012}. One interpretation of this low-temperature nonlinear response is that it is due to an intrinsic nonlinearity arising from Josephson coupling between two or more gaps in the Ba(Fe$_{1-x}$Co$_x$)$_2$As$_2$ film. Hence, the variable phase difference of the coupled superconducting gaps will produce the nonlinear response when a nonequilibrium charge imbalance appears at short length scales \cite{Gurevich}, essentially the nonlinear mechanism associated with excitation of the Leggett mode for two gap superconductors \cite{Gurevich}. Note that the situation of charge imbalance can be easily created from the perpendicular component of the RF magnetic field as shown in Fig. \ref{yoke}.

Secondly, after a little increase of $P_f$ from the point of turnover on each $P_{3f}$ versus $P_f$ curve, one can see a sharp drop at T=5 K and T=10 K in Fig. \ref{P3fPf} (a). These sharp drops suggest the sudden loss of a nonlinearity mechanism.  For example it could be the annihilation of a superconducting order parameter in high RF magnetic field, implying that perhaps one of the superconducting gaps is suddenly destroyed by localized intense magnetic field on the Ba(Fe$_{1-x}$Co$_x$)$_2$As$_2$ film surface. At that point, the Leggett mode nonlinearity is eliminated due to the loss of the second gap, reducing the $P_{3f}$ output suddenly.

\begin{figure}
    \centering
    \includegraphics[width=2.5in]{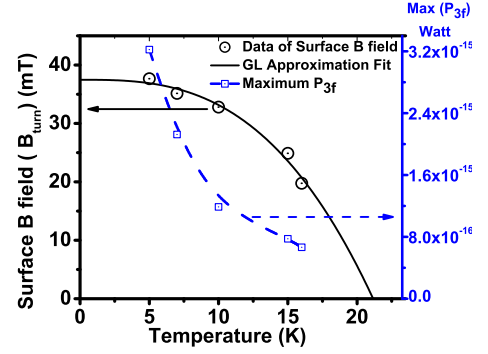}
    \caption{Temperature dependent estimated surface critical magnetic field of Ba(Fe$_{1-x}$Co$_x$)$_2$As$_2$ and $P_{3f}$ in Watts at the turnover point. Note that the dashed line is a B-Spline fit. The solid line is a fit of the surface critical magnetic field from Eq. \ref{approximation}. with $T_c'=21.15$ $K$, $n=2.88$ and $B_0=37.47$ $mT$.}
    \label{Ba122_critical_power}
\squeezeup
\end{figure}

The excitation level of these turnover points should be proportional to the temperature dependent lower critical field of the smallest energy gap of Ba(Fe$_{1-x}$Co$_x$)$_2$As$_2$. Fig. \ref{Ba122_critical_power} shows a summary plot of the $P_{3f}$ and estimated surface B field at the turnover peak points ($B_{turn}$) versus the corresponding temperature. Note that $P_{3f}$ is on a linear power scale in Watts. The surface RF magnetic field is converted from $P_f$ by the the relation of $P_{f}(T)=k[B_{turn}(T)]^2$, where  $\textit{k}$ a constant relating the incident power in the probe to the RF magnetic field experienced by the sample surface, and $k=25.6 W/T^2$ is taken \cite{Tai2014_APL}. This number is judged by the field scale generated by HFSS simulation in Fig. \ref{setup}(c) and experimental results on a known Nb conventional superconductor for calibration.
One can find the temperature dependent $P_{3f}$ and surface $B$ field at the turnover peak points is similar to the temperature dependent critical field. A fit of temperature dependent surface $B_{turn}$ field is done by tuning the value of $T_c'$, $n$ and $B_0$ of the following approximation equation:
\begin{equation}\label{approximation}
B_{turn}(T)=B_0(1-(\frac{T}{T_c'})^n)
\squeezeup
\end{equation}
With $T_c'=21.15$ $K$, $n=2.88$ and $B_0=37.47$ $mT$, one can get the smallest standard deviation between experimental data points and the approximation equation as shown on the plot of the solid line in Fig. \ref{Ba122_critical_power}. The $B_0$ perhaps can be interpreted as the lower critical field $B_{c1}$ for fields along the c-axis of the Ba-122 film. Given an upper critical field $B_{c2}(T=0 K)$$\sim$100 $T$ and Ginzburg-Landau (GL) parameter $\kappa$=$\lambda/\xi_{GL}$ $\sim$ $10^2$ \cite{Gurevich_2011}\cite{Putti}, where $\xi_{GL}$ is the superconducting coherence length, one can predict the value of $B_{c1}$ for Ba-122 single crystal is on the order of 25 $mT$ \cite{Gurevich_2011}. The $B_0$ from the fitting is also very close to the experimentally measured lower critical field on another iron-pnictide superconductor $BaFe_2(As_{1-x}P_x)_2$ with $B_{c1}(T=0 K)$= 30 mT $\sim$ 60 mT at different Phosphorus doping \cite{Putzke}.

Finally, for Ba(Fe$_{1-x}$Co$_x$)$_2$As$_2$ films as shown in Fig. \ref{P3fPf}(a), the shape of $P_{3f}$ on $P_{f}$ at T=16.5 K is different from that at T=5 K and 10 K. At T=16.5 K, there is no sharp drop after the turnover point. This shape is similar to that of bulk Nb at any selected temperature. This implies that at T=16.5 K, nonlinearity comes from the extrinsic nonlinear response, and something analogous is seen in bulk Nb. This specific curve shape is very similar to the interference from vortices in Josephson coupled rings \cite{Tai2014_APL}\cite{Tai_2015arXiv}, which indicates the nonlinearity at T=16.5 K comes from the dynamics of vortex penetration, jumpwise instabilities in the Ba(Fe$_{1-x}$Co$_x$)$_2$As$_2$ film under strong and localized RF fields. An extrapolation dashed line similar to the plot on Nb may indicate the presence of another superconducting gap (large gap) of Ba(Fe$_{1-x}$Co$_x$)$_2$As$_2$ at this temperature. Hence the co-existence of a sharp drop of $P_{3f}$ on $P_f$ and extrapolation plot to indicate the annihilation of superconducting gaps on the $P_{3f}$ versus $P_f$ curves suggests a multigap nature of Ba(Fe$_{1-x}$Co$_x$)$_2$As$_2$.

\section{Summary}
From the temperature dependent linear responses measurement, the Ba(Fe$_{1-x}$Co$_x$)$_2$As$_2$ film shows a relatively wide superconducting transition width at its $T_c$. From the temperature dependent $P_{3f}(T)$ measured at 8 dBm excitation power, different nonlinear mechanisms in different temperature regions are seen. At temperatures around $T_c$, the nonlinearity comes from the current-induced modulation of the suppressed superconducting order parameter near $T_c$. In the intermediate temperature range, nonlinearity is dominated by the motion of vortices. At temperatures below 6 K, in addition to the possible NS junction nonlinearity, the measured nonlinearity of the Ba(Fe$_{1-x}$Co$_x$)$_2$As$_2$ film is also consistent with the intrinsic nonlinear Meissner Effect of a nodal gap function at low temperature. The third harmonic power ($P_{3f}$) dependence on fundamental tone power ($P_f$) shows nonlinearity at low temperature can be easily stimulated at very low excitation power, quite different from the results on Nb, a conventional single gap s-wave superconductor. Therefore, from the localized high field electrodynamic measurements, the Ba(Fe$_{1-x}$Co$_x$)$_2$As$_2$ superconductor shows behavior consistent with a multigap nature and possibly a nodal gap.

\section{Acknowledgement}
This work is supported by the US Department of Energy $/$ High Energy Physics through grant $\#$ DE-SC0012036T and CNAM. Research at UW-Madison (Design and synthesis of thin film heterostructures used in this work) was supported
by the US Department of Energy, Office of Basic Energy Sciences, Division of Materials Sciences and
Engineering under award number DE-FG02-06ER46327 (C.B.E.). \\



\begin{thebibliography}{99}

\bibitem{Paglione}
J. Paglione, R. L. Greene ``High-temperature Superconductivity in
Iron-based Materials," Nature Physics, \textbf{6}, p. 645, (2010).

\bibitem{Gurevich_2011}
A. Gurevich, ``Iron-based Superconductors at High Magnetic Fields," Rep. Prog. Phys. \textbf{74}, p. 124501, (2011).

\bibitem{Terashima}
K. Terashima, Y. Sekiba, J. H. Bowen, K. Nakayama, T. Kawahara, T. Sato, P. Richard, Y.-M. Xu, L. J. Li, G. H. Cao, Z.-A. Xu, H. Ding, T. Takahashi,
``Fermi Surface Nesting Induced Strong Pairing
in Iron-based Superconductors," Proceedings of the National Academy of Sciences (PNAS), \textbf{106}, no. 18, p. 7330 (2009).

\bibitem{S. Lee_2010}
S. Lee and and J. Jiang, Y. Zhang and C. W. Bark and J. D. Weiss and C. Tarantini and C. T. Nelson and H. W. Jang and C. M. Folkman and S. H. Baek and A. Polyanskii and D. Abraimov and A Yamamoto and J. W. Park and X. Q. Pan and E. E. Hellstrom and D. C. Larbelestier and C. B. Eom, ``The Behavior of Grain Boundaries in the Fe-based Superconductors," Nature Materials \textbf{9}, p. 397, (2010).

\bibitem{S. Lee_2011}
S. Lee, J. Jiang, J. D. Weiss, C. W. Bark, C. Tarantini, M. D. Biegalski, A. Polyanskii, Y. Zhang, C. T. Nelson, X. Q. Pan, E. E. Hellstrom, D. C. Larbalestier, C. B. Eom, ``Dependence of Epitaxial Ba(Fe$_{1-x}$Co$_x$)$_2$As$_2$ Thin Films Properties on SrTiO$_3$ Template Thickness," IEEE Trans. Appl. Supercond. \textbf{21}, No. 3, p. 2882, (2011).

\bibitem{Chen}
T. Y. Chen, Z. Tesanovic, R. H. Liu, X. H. Chen, C. L. Chien. "A BCS-like Gap in the Superconductor SmFeAsO$_{0.85}$F$_{0.15}$," Nature (London) \textbf{453}, p. 1224, (2008).

\bibitem{Dressel}
B. Gorshunov, D. Wu, A. A. Voronkov, P. Kallina, K. Iida, S. Haindl, F. Kurth, L. Schultz, B. Holzapfel, M. Dressel, ``Direct Observation of the Superconducting Energy Gap in the Optical Conductivity
of the Iron Pnictide Superconductor Ba(Fe$_0.9$Co$_0.1$)$_2$As$_2$," Phys. Rev. B \textbf{81}, p. 060509(R), (2010).

\bibitem{Perucchi}
A. Perucchi, L. Baldassarre, S. Lupi, J. Y. Jaing, J. D. Weiss, E. E. Hellstrom, S. Lee, C. W. Bark, C. B. Eom, M. Putti, I. Pallecchi, C. Marini, P. Dore, ``Multi-gap Superconductivity in a $BaFe_{1.84}Co_{0.16}As_2$ Film From Optical Measurements at Terahertz Frequencies," Eur. Phys. J. B \textbf{77} p. 25-30, (2010).

\bibitem{Yong}
J. Yong, S. Lee, J. Jiang, C. W. Barj, J. D. Weiss, E. E. Hellstrom, D. C. Larbalestier, C. B. Eom, T. R. Lemberger, ``Superfluid Density Measurements of Ba(Fe$_{1-x}$Co$_x$)$_2$As$_2$ Films Near Optimal Doping," \textbf{83}, Phys. Rev. B  P. 104510, (2011).

\bibitem{Xiaohang}
Xiaohang Zhang, Yoon Seok Oh, Yong Liu, Liqin Yan, Shanta R. Saha, Nicholas P. Butch, Kevin Kirshenbaum, Kee Hoon Kim, Johnpierre Paglione, Richard L. Greene, Ichiro Takeuchi, ``Evidence of a Universal and Isotropic 2$\Delta$/k$_{B}$T$_{c}$ Ratio in 122-type Iron Pnictide Superconductors Over a Wide Doping Range," Phys. Rew. B \textbf{82}, p. 020515(R), (2010).

\bibitem{Prozorov}
R. T. Gordon, N. Ni, C. Martin, M. A. Tanatar, M. D. Vannette, H. Kim, G. D. Samolyuk, J. Schmalian, S. Nandi, A. Kreyssig, A. I. Goldman, J. Q. Yan, S. L. Bud'ko, P. C. Canfield, R. Prozorov, ``Unconventional London Penetration Depth in Single-Crystal Ba(Fe$_{1-x}$Co$_x$)$_2$As$_2$ Superconductors," Phys. Rev. Lett. \textbf{102}, P. 127004, (2009).

\bibitem{Fisher}
B. Muschler, W. Prestel, R. Hackl, T. P. Devereaux, J. G. Analytis, Jiun-Haw Chu, I. R. Fisher, ``Band- and Momentum-dependent Electron Dynamics in Superconducting Ba(Fe$_{1-x}$Co$_x$)$_2$As$_2$ as Seen via Electronic Ramman Scattering," \textbf{80}, p. 180510(R), (2009).

\bibitem{Moler2009}
Clifford W. Hicks and Thomas M. Lippman and Martin E. Huber and James G. Analytis and Jiun-Haw Chu and Ann S. Erickson,
Ian R. Fisher and Kathryn A. Moler, ``Evidence for a Nodal Energy Gap in the Iron-Pnictide Superconductor LaFePO from Penetration Depth Measurements by Scanning SQUID Susceptometry," Phys. Rev. Lett. \textbf{103}, p. 127003, (2009)

\bibitem{Moler2010}
B. Kalisky and John R. Kirtley and J. G. Analytis and Jiun-Haw Chu and A. Vailionis and Ian R. Fisher, Kathryn A. Moler, ``Stripes of Increased Diamagnetic Susceptibility in Underdoped Superconducting $Ba(Fe_{1-x}Co_{x})_2As_2$ Single Crystals:
Evidence for an Enhanced Superfluid Density at Twin Boundaries," Phys. Rev. B \textbf{81}, p. 184513, (2010)

\bibitem{Larbalestier_2011}
J. H. Durrell and C-B Eom and E. E. Hellstrom and C. Tarantini, A. Yamamoto, D. C. Larbelestier,``The Behavior of Grain Boundaries in the Fe-based Superconductors," Reports on Progress in Physics \textbf{74}, p. 124511, (2011).

\bibitem{Tarantini}
C. Tarantini, S Lee, Y Zhang, J. Jiang, C. W. Bark, J. D. Weiss, A. Polyanskii, C. T. Nelson, H. W. Jang, C. M. Folkman, S. H. Baek, X. Q. Pan, A. Gurevich, E. E. Hellstrom, C. B, Eom, D. C. Larbalestier,``Strong Vortex Pinning in Co-doped BaFe$_2$As$_2$ Single Crystal Thin Film," Appl. Phys. Lett. \textbf{96}, p. 142510, (2010).

\bibitem{Zhang}
Y. Zhang, Christopher T. Nelson, Sanghan Lee, Jianyi Jiang, Chung Wung Bark, Jeremy D. Weiss, Chiara Tarantini, Chad M. Folkman, Seung-Hyub Baek, Eric E. Hellstrom, David C. Larbalestier, Chang-Beom Eom, Xiaoqing Pan, ``Self-assembled Oxide Nanopillars in Epitaxial BaFe$_2$As$_2$ Thin Films for Vortex Pinning," Appl. Phys. Lett. \textbf{98}, p. 042509, (2011).

\bibitem{Anlage_2001}
Steven M. Anlage, D. E. Steinhauer, B. J. Feenstra, C. P. Vlahacos, and F. C. Wellstood, ``Near-Field Microwave Microscopy of Materials Properties," in Microwave Superconductivity, ed. by H. Weinstock and M. Nisenoff, (Kluwer, Amsterdam, 2001), p. 239-269.

\bibitem{Anlage_2007}
Steven M. Anlage, Vladimir V. Talanov, Andrew R. Schwartz, ``Principles of Near-Field Microwave Microscopy in Scanning Probe Microscopy: Electrical and Electromechanical Phenomena at the Nanoscale," Volume 1, edited by S. V. Kalinin and A. Gruverman (Springer-Verlag, New York, 2007, ISBN: 978-0-387-28667-9), p. 215-253.

\bibitem{Tai2011}
Tamin Tai, X. X. Xi, C. G. Zhuang, D. I. Mircea, S. M. Anlage, ``Nonlinear Near-Field Microwave Microscope for RF Defect Localization in Superconductors,¡¨ IEEE Trans. Appl. Supercond.
\textbf{21}, 2615, (2011).

\bibitem{Tai2012}
Tamin Tai, B. G. Ghamsari, T. Tan, C. G. Zhuang, X. X. Xi, Steven M. Anlage, ``MgB$_2$ Nonlinear Properties Investigated Under Localized High RF Magnetic Field Excitation," Phys. Rev. ST Accel. Beams \textbf{15}, 122002, (2012).

\bibitem{Tai2013}
Tamin Tai, B. G. Ghamsari, Steven M. Anlage, IEEE Trans. ``Nanoscale Electrodynamic
Response of Nb Superconductors,¡¨ Appl. Supercond. \textbf{23}, 7100104, (2013).

\bibitem{Tai2014_JAP}
Tamin Tai, B. G. Ghamsari, Steven M. Anlage, ``Modeling the Nanoscale Linear Response of Superconducting Thin Films Measured by a Scanning Probe Microwave Microscope," Journal of Appl. Phys.  \textbf{115}, 203908, (2014).

\bibitem{Tai2014_APL}
Tamin Tai, B. G. Ghamsari, Tom Bieler, Steven M. Anlage, ``Nanoscale Nonlinear Radio Frequency Properties of Bulk Nb: Origins of Extrinsic Nonlinear Effects," Appl. Phys. Lett. \textbf{104}, 232603, (2014).

\bibitem{Tai2013_dissertation}
Tamin Tai, Ph.D. dissertation, University of Maryland-College Park, (2013),
see http://hdl.handle.net/1903/14668.

\bibitem{Lee}
S. C. Lee, M. Sullivan, G. R. Ruchti, and S. M. Anlage, ``Doping
Dependent Nonlinear Meissner Effect and Spontaneous Currents in
High-$T_c$ Superconductors," Phys. Rev. B \textbf{71}, p.
014507, 2005.

\bibitem{Mircea}
D. I. Mircea, H. Xu, S. M. Anlage, ``Phase-sensitive Harmonic
Measurements of Microwave Nonlinearities in Cuprate Thin Films,"
Phys. Rev. B \textbf{80}, p. 144505, (2009).

\bibitem{Jeffries}
C. Jeffries, Q. H. Lam, Y. Kim, L. C. Bourne, A. Zettl, ``Symmetry Breaking and Nonlinear Electrodynamics in the Ceramic Superconductor YBa$_2$Cu$_3$O$_7$," Phys. Rev. B \textbf{37}, p. 9840, (1988).

\bibitem{Lee2}
S. C. Lee, S.Y. Lee, S. M. Anlage, ``Microwave Nonlinearities of an Isolated Long YBa$_2$Cu$_3$O$_{7-\delta}$ Bicrystal Grain Boundary," Phys. Rev. B \textbf{72}, p. 024527, (2005).

\bibitem{L. Ji}
L. Ji, R. H. Sohn, G. C. Spalding, C. J. Lobb, and M. Tinkham, ``Critical State Model for Harmonic Generation in High-temperature Superconductors,¡¨ Phys. Rev. B \textbf{40}, p. 10936-10945, (1989).

\bibitem{Gurevich2008}
A. Gurevich, G. Ciovati, ¡§Dynamics of vortex penetration, jumpwise instabilities,
and nonlinear surface resistance of type-II superconductors in strong rf
fields,¡¨ Phys. Rev. B 77, 104501, (2008).

\bibitem{Golosovsky}
M. Golosovsky, D. Davidov, E. Farber, T. Tsach, M. Schieber, ``Microwave Transmission and Harmonic Generation in Granular High-Tc Superconducting Films: Evidence for Viscous Flux Motion and Weak Links," Phys. Rev. B \textbf{43}, p. 10390 (1991).

\bibitem{E. B. Sonin}
E. B. Sonin and A. K. Tagantsev, ``Vortex Convection Non-linearity of AC Response of Superconductors with Weak Links," Supercond. Sci. Tech. \textbf{4}, p. 119-121 (1991).

\bibitem{M. W. Coffey}
M. W. Coffey, ``Coupled Nonlinear Electrodynamics of Type-II Superconductors in the Mixed State," Phys. Rev. B \textbf{46}, p. 567-570 (1992).

\bibitem{oates}
D. E. Oates, S.-H. Park, G. Koren, ``Observation of the Nonlinear Meissner Effect in YBCO Thin Films: Evidence for a d-Wave
Order Parameter in the Bulk of the Cuprate Superconductors,"  Phys. Rev. Lett.
\textbf{93}, p. 197001 (2004).

\bibitem{Agassi}
D. Agassi, D. E. Oates, ``Nonlinear Meissner Effect in a High-temperature Superconductor,"
Phys. Rev. B \textbf{72}, p. 014538 (2005).

\bibitem{oates_2010}
D. E. Oates, Y. D. Agassi, B. H. Moeckly, ``Microwave measurements of MgB2: implications for applications and order-parameter symmetry," Supercond. Sci. Technol. \textbf{23}, 034011, (2010).

\bibitem{Lesovik}
G. B. Lesovik, A. L. Fauch$\grave{e}$re ,G. Blatter, ``Nonlinearity in Normal-metal-superconductor Transport: Scatterring-matrix Approach," Phys. Rev. B \textbf{55}, p. 3146 (1997).

\bibitem{Gurevich}
A. Gurevich, V. M. Vinokur,``Interband Phase Modes and Nonequilibrium
Soliton Structures in Two-Gap Superconductors,"  Phys. Rev. Lett.  \textbf{90}, p. 047004 (2003).

\bibitem{Putti}
M. Putti, I. Pallecchi, E. Bellingeri, M. R. Cimberle, M. Tropeano,
C. Ferdeghini, A. Palenzona, C. Tarantini, A. Yamamoto, J. Jiang,
J. Jaroszynski, F. Kametani, D. Abraimov, A. Polyanskii,
J. D. Weiss, E. E. Hellstrom, A. Gurevich, D. C. Larbalestier, R. Jin,
B. C. Sales, A. S. Sefat, M. A. McGuire, D. Mandrus, P. Cheng,
Y. Jia, H. H. Wen, S. Lee and C. B. Eom, ``New Fe-based Superconductors: Properties
Relevant for Applications," Supercond. Sci. Technol.,
\textbf{23}, p. 034003, (2010).

\bibitem{Putzke}
C. Putzke, P. Walmsley, J. D. Fletcher, L. Malone, D. Vignolles, C. Proust, S. Badoux, P. See, H. E. Beere, D. A. Ritchie, S. Kasahara, Y. Mizukami, T. Shibauchi, Y. Matsuda, A. Carrington,``Anomalous critical fields in Quantum Critical Superconductors," Nature Communication, \textbf{5}, p. 5679, (2014).

\bibitem{Tai_2015arXiv}
T. Tai, B. G. Ghamsari, T. Bieler, S. M. Anlage, ``Nanoscale Nonlinear Radio Frequency Properties of Bulk Nb : Origins of Extrinsic Nonlinear Effects," arXiv:1507.02777

\end{thebibliography}
\end{document}